# Bi-directional ultrafast electric-field gating of interlayer transport in a cuprate superconductor

A. Dienst[(1)], M. Hoffmann[(2)], D. Fausti [(1,2)], J. Petersen[(1,2)], S. Pyon[(3)], T. Takayama[(3)], H. Takagi[(3,4)], A. Cavalleri[(1,2)]

[(1)] Department of Physics, Clarendon Laboratory, University of Oxford, United Kingdom
[(2)] Max Planck Research Department for Structural Dynamics, University of Hamburg-CFEL
[(3)] Department of Advanced Materials Science, University of Tokyo, Tokyo, Japan
[(4)] RIKEN Advanced Science Institute, Hirosawa 2-1, Wako 351-0198, Japan

## Abstract

In cuprate superconductors, tunneling between planes makes possible three-dimensional coherent transport. However, the interlayer tunnelling amplitude is reduced when an order-parameter-phase gradient between planes is established. As such, c-axis superconductivity can be weakened if a strong electric field is applied along the c axis. We use high-field single-cycle terahertz pulses to gate interlayer coupling in $La_{1.84}Sr_{0.16}CuO_4$. We induce ultrafast oscillations between superconducting and resistive states and switch the plasmon response on and off, without reducing the density of Cooper pairs. Indeed, in-plane superconductivity remains unperturbed throughout, revealing a non-equilibrium state in which the dimensionality of the superconductor is time dependent. The gating frequency is determined by the electric field strength, in the spirit of the ac Josephson effect. Non-dissipative, bi-directional gating of superconductive coupling is of interest for device applications in ultrafast nanoelectronics. It is also a new example of nonlinear terahertz physics, applicable to nanoplasmonics and active metamaterials.

Transport in cuprate superconductors can be understood by considering a stack of intrinsic Josephson junctions, made of superconducting planes separated by insulating layers. Three key features characterize the c-axis electrodynamics in the superconducting state. First, the dc resistivity vanishes, as superconductive tunneling "shorts" resistive transport through incoherent quasi-particles. Second, the imaginary part of the conductivity displays $1/\omega$ frequency dependence, reflecting diamagnetism and the Meissner effect as $\omega \to 0$. Third, the combination of tunneling, which has an equivalent inductive impedance, and capacitive coupling between the planes, leads to collective plasma oscillations of superconducting electrons at terahertz (THz) frequencies, or Josephson plasma waves *(1)*.

These properties, typically measured in the frequency-domain *(2)*, can be observed directly with time-domain THz spectroscopy *(3)*. Figure 1A shows one such electro-optic-sampling measurement of a single-cycle THz-field, after reflection off the optimally-doped cuprate $La_{1.84}Sr_{0.16}CuO_4$. In the superconducting phase (red curve), long-lived oscillations at 2-THz frequency appear on the trailing edge of the pulse. The incident field was measured after reflection from a gold-coated fraction of the sample surface. The frequency-dependent complex reflection coefficient $r(\omega) = E_{refl}(\omega)/E_{inc}(\omega)$, was then derived by dividing the Fourier transforms of the time dependent reflected field of figure 1 by the incident one. The reflectivity $|r(\omega)|$ is displayed in Fig. 1B, and reproduces well the Josephson plasma edge in this compound.

In figure 1C, the complex frequency dependent dielectric function $\varepsilon(\omega)$ of the equilibrium low temperature state is displayed, obtained by fitting the reflectivity $|r(\omega)|$ with the two-fluid model *(4)*. As for any plasmonic response, the real part $Re\{\varepsilon(\omega)\}$ is negative for $\omega < \omega_p$, where $\omega_p/2\pi = 2$ THz is the frequency of the Josephson plasma resonance in $La_{1.84}Sr_{0.16}CuO_4$. The imaginary part $Im\{\varepsilon(\omega)\}$ is nearly zero over the whole frequency range, indicating negligible dissipation by non-superconducting quasi-particles. In figure 1D, we also display the low temperature conductivity $\sigma(\omega)$, characterized by vanishing real part for finite frequencies, and a $1/\omega$ frequency dependence in its imaginary component, $Im\{\sigma(\omega)\} = \rho/4\pi\omega$. Here, $\rho$ is the superfluid density, a measure for the stiffness of the condensate at equilibrium *(5)*. These equilibrium transport properties have been discussed extensively in the past, especially in regard to the controversial role of interlayer tunneling in high-$T_c$ superconductivity *(6,7,8)*.

The present work aims at perturbing coherent Josephson coupling without injecting incoherent excitations, to gate c-axis transport at high repetition rates. It is known that interlayer transport can be altered statically by application of magnetic *(9)* or electric fields *(10)*. This is possible because tunneling across a weak link depends on the order-parameter-phase difference between the two superconductors ($\phi$), which is affected by application of external fields *(11,12)*. To achieve this effect on the ultrafast timescale, interlayer voltage drops of few to tens of mV are needed, corresponding to THz-frequency transients with peak electric fields of tens of kV/cm.

High-field THz transients were achieved in our experiments with the tilted-pulse front technique, which was used to generate μJ single-cycle pulses by optical rectification in LiNbO$_3$ *(13)*. These pulses were tuned to a centre frequency of 450 GHz, well below the 2 THz Josephson plasma edge (see Fig. 2). The gate field wavelength was ~ 0.65 mm, and could be focused down to spot sizes of approximately 1 mm$^2$ to reach field strengths up to 100 kV/cm. The gate field was polarized perpendicular to the planes and was completely reflected, penetrating over a distance of 5 μm as an evanescent wave.

The time dependent reflectivity of La$_{1.84}$Sr$_{0.16}$CuO$_4$ was probed both perpendicular and parallel to the planes with a delayed THz probe pulse. The probe bandwidth extended up to 2.5 THz and, for c-axis polarization, it covered the Josephson plasma edge (see Fig. 2). To extract the time dependent conductivity, the amplitude- and phase-resolved transients were fitted by a model that considered a 5-μm-thick surface layer of unknown conductivity, over an unperturbed semi-infinite superconductor with the optical properties of Fig. 1.

The key observation of our work is reported in the two-dimensional plots of figure 3A, which display the frequency dependent conductivities (real and imaginary part) for different time delays $\tau$ between an 80 kV/cm single-cycle gate field and the probe pulse. In the two upper panels, c-axis measurements are displayed. As the gate electric field evolves in time, superconductive coupling vanishes for $\tau = 1.25$ ps, as qualitatively shown by the loss of spectral weight in the imaginary conductivity, and in the corresponding gain of the real part. Remarkably, superconductive transport is re-established within a few hundred femtoseconds ($\tau = 1.5$ ps), when the conductivity of the unperturbed superconductor is recovered. Oscillations between the two states follow. Fig. 3B shows lineouts of the complex conductivity at the peak and the troughs of these oscillations. At negative time delays and in the recurring superconducting states, *Re{$\sigma(\omega,\tau<0)$}* nearly vanishes at all frequencies, whilst *Im{$\sigma(\omega,\tau<0)$}* follows a *$1/\omega$* frequency dependence, as in figure 1D. At time delays were resistive states are established (dashed curve), the real conductivity is the dominant contribution and tends to a finite value $\sigma_0$ for $\omega \to 0$, as expected for a Drude gas of incoherent quasi-particles. In this state, the imaginary conductivity still exhibits its *$1/\omega$* dependence, but with a strongly depleted pre-factor.

A second important observation results from the *ab*-plane conductivity, which remains essentially unperturbed throughout these dynamics as displayed in the lower panels of figure 3A. Effectively, the dimensionality of the cuprate superconductor oscillates in time as the planes are decoupled, an exotic phenomenon never observed to date. The fact that the in-plane optical properties do not show significant change reinforces the notion that the THz field perturbs the phase but does not ionize the Cooper pairs, effectively maintaining the modulus of the order parameter unperturbed.

The physics observed here can be quantitatively discussed as follows. The interlayer coupling strength can be described as an equivalent tunneling inductance $L_s$, proportional to $1/\cos\phi$ *(12)*. At equilibrium $\phi \approx 0$, the inductance $L_s$ is minimum and transport by non-condensed, incoherent quasi-particles is optimally "shorted". For weak electric fields, a supercurrent $J_s$ is driven through the layers, and a gradient in $\phi$ develops across the junction, as, according to the first Josephson equation, $J_s \propto \sin\phi$. For large fields, as $L_s$ increases, ohmic conduction by quasi-particles becomes relevant and a voltage drop $V$ develops across the junction.

At this point the order parameter phase difference starts advancing in time according to the second Josephson equation, i.e. as $\dot\phi = 2eV(t)/\hbar$, where $e$ is the electron charge and $\hbar$ Planck's constant. The tunneling evolves then as $L_s \propto 1/\cos\phi$, diverging when $\phi$ crosses $\pm\pi/2$ and changing sign in between.

In the upper panel in figure 4A we report the normalized time integral of the gate electric field $\int E_{pump}(t)dt$, derived from the measured THz pump transients $E_{pump}(t)$. The integral of the field is compared to the measured time dependent strength of c-axis superconducting transport, evaluated as $S(\tau) = \lim_{\omega \to 0} \omega \, \text{Im}(\sigma(\omega,\tau))$. At equilibrium, this quantity is proportional to the superfluid density $\rho$. Here, $S(\tau)$ is used as a measure of the non-equilibrium interlayer coupling strength and is well fitted by the function $|\cos(c\int E_{pump}(t)dt)|$ with only the constant $c$ left as free parameter. As the integral of the driving voltage evolves, and as $\phi$ crosses $\pi/2$, $S(\tau)$ vanishes, recovering as $\phi$ is driven towards $\pi$. The model does not fit well at low field strengths, as, according to what discussed above, the transport is still superconducting and no voltage drop develops. At high fields, the model is instead faithfully reproducing the measured non-equilibrium interlayer coupling strength $S(\tau)$ over a large dynamic range.

The superconductive-resistive oscillation frequency, measured by sampling the peak of the probe THz transient as function of delay, is shown in Fig. 4B as a function of pump field strength. Two regions are identified. Below $E^* = 75$ kV/cm, the transport is modulated at the frequency of the pump. In this regime the interlayer phase difference is perturbed but it does not reach $\pi/2$ and the coupling is never completely shut off. For field strengths above $E^*$, the modulation frequency increases linearly with the field in spirit of the ac Josephson effect.

This voltage to frequency conversion is a new demonstration of nonlinear THz physics that could be extended to photonic devices and modulators, but also to nanoplasmonic devices. Secondly, we comment on the potential applications to electronics. Because interlayer transport is determined by short range tunneling between neighboring layers, this effect could be applied to single nanoscale junctions. Finally, strong field perturbations of interlayer couplings *(14)* may be used to test new ideas of the physics of cuprates, including the case of striped states for which Josephson de-coupling may be important *(14)*.

**FIGURE CAPTIONS**

**Fig1: (A)** Reflected THz electric fields from $La_{1.84}Sr_{0.16}CuO_4$, above (black curve, 40 K) and below (red curve, 5 K) $T_c$ = 36 K. The probe electric field is polarized along the c axis, probing interlayer transport. **(B)** Electric field amplitude reflectivity spectrum above (black curve, 40 K) and below (red curve, 5 K) $T_c$ = 36 K. **(C)** Real and imaginary part of the frequency dependent equilibrium permittivity $\varepsilon(\omega)$ at 5 K. While $Im\{\varepsilon(\omega)\}$ is small indicating negligible dissipation, $Re\{\varepsilon(\omega)\}$ highlights the plasmonic response crossing zero at the plasma frequency $\omega_p/2\pi$ = 2 THz. **(D)** Real and imaginary part of the equilibrium conductivity $\sigma(\omega)$ in the superconducting state at 5 K. $Re\{\sigma(\omega)\}$ is small and almost frequency independent since the spectral weight is accumulated in the superconducting delta function at zero frequency. $Im\{\sigma(\omega)\}$ follows $1/\omega$ representing the Meissner effect and diamagnetic response at low frequencies.

**Fig2:** Normalized spectra of THz pump (thick black curve) and probe (thin black curve) pulses, overlaid to the $La_{1.84}Sr_{0.16}CuO_4$ reflectivity edge at 5K. The inset shows typical time domain transients of THz pump and probe pulses measured by electro-optic sampling.

**Fig3: (A)** The upper panels show two-dimensional plots of real and imaginary part of the c-axis conductivity $\sigma_c(\omega,\tau)$ for different pump probe time delays, measured at 5 K. Pump and probe polarisation are parallel to the c-axis. The pump field strength is 80 kV/cm. For $\tau < 0$, the line-outs of figure 3A reflect the unperturbed optical conductivity. As the gate field progresses, the conductivity oscillates rapidly between resistive and superconductive state. Oscillations persist for a couple of picoseconds, comparable to the Josephson coherence time. The lower two panels display the low temperature in-plane conductivity $\sigma_{ab}(\omega,\tau)$, measured by rotating the probe polarization by 90 degree to be parallel to the ab-plane, whilst the pump polarization is kept parallel to the c-axis. **(B)** Lineouts of conductivity $\sigma_c(\omega,\tau)$ at the peak and the troughs of the oscillations, exhibiting resistive (1.25 ps) and superconductive (1.5 ps) coupling.

**Fig4: (A)** The upper panel shows the normalized integral of the measured THz pump transient as a function of time delay, being proportional to the advancement of the interlayer phase difference $\phi(t) \propto \int^{t} E_{pump}(t')dt'$. The lower panel shows $S(\tau) = \lim_{\omega \to 0} \omega \, \text{Im}(\sigma(\omega,\tau))$ as a measure of interlayer coupling strength extracted from our experimental data (red dots). The data points are fit by $|\cos(c\int E_{pump}(t)dt)|$ (black dashed line), where $c$ is left as free parameter. **(B)** Modulation frequency in dependence of THz gate electric field strength. Two regions can be distinguished: for low amplitudes $E_{pump} < E^{*}$, the modulation occurs at the pump frequency (the shaded area represents the full width at half maximum of the probe amplitude spectra). Above the critical threshold $E^{*}$, the modulation frequency increases linearly with field strength, reminiscent of the ac Josephson effect.

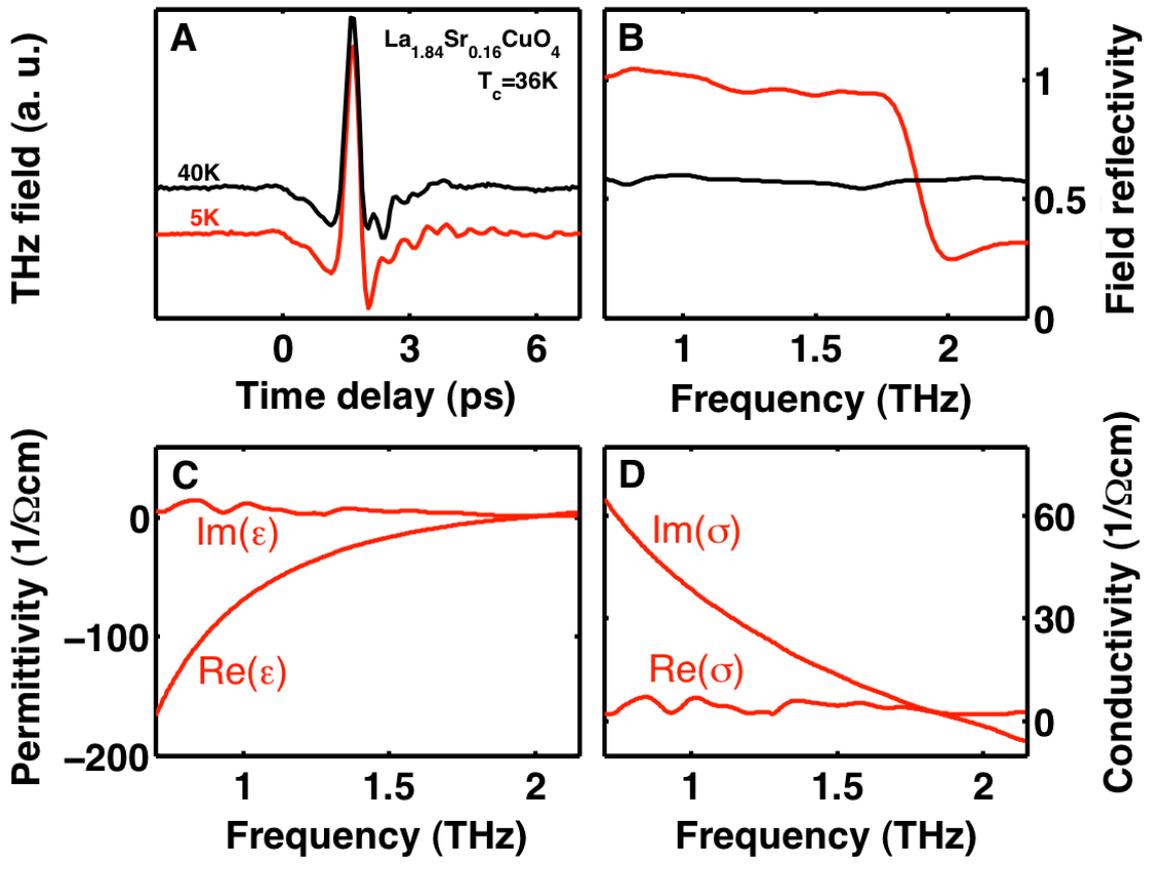

**Figure 1**

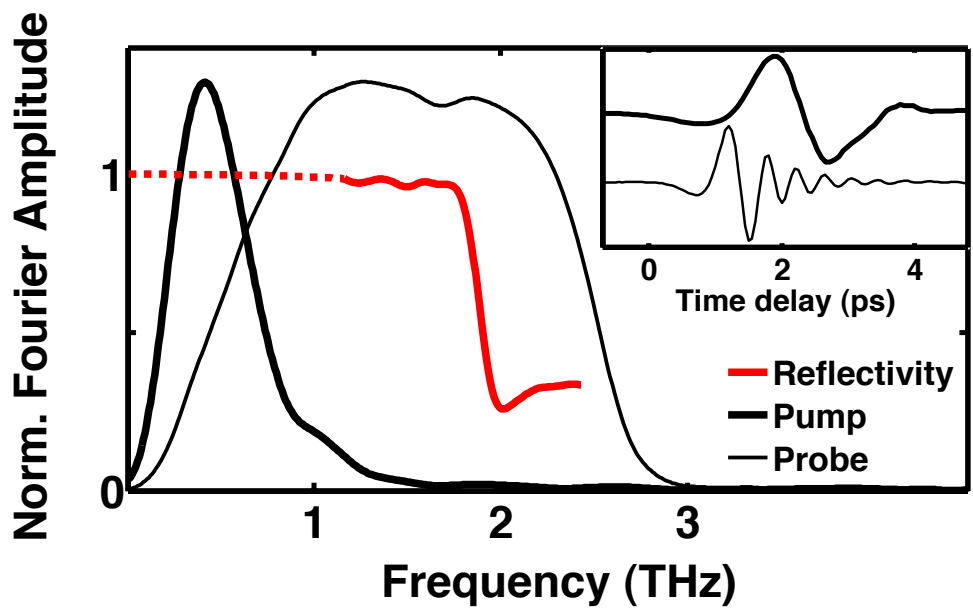

**Figure 2**

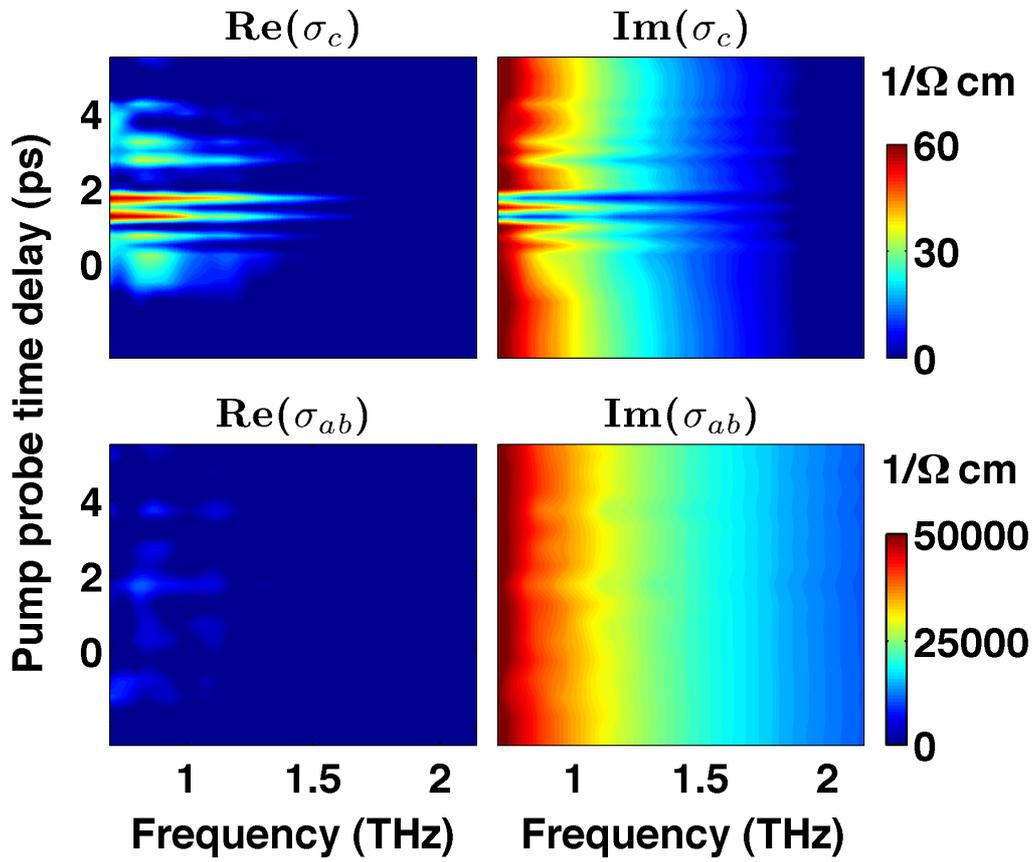

Figure 3A

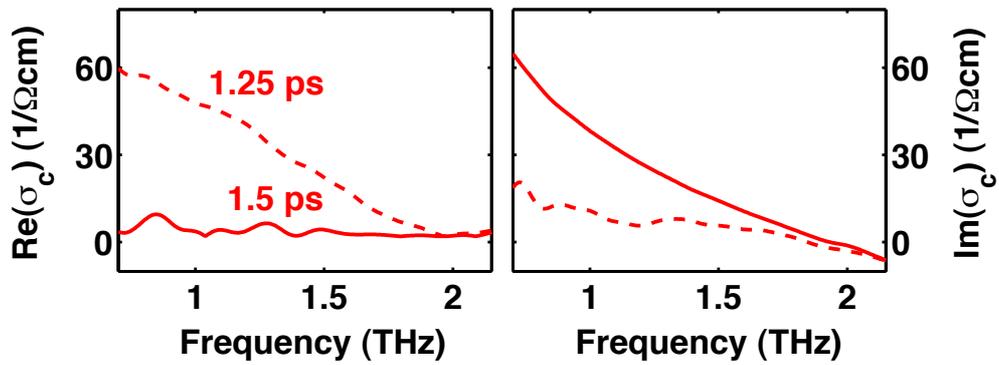

Figure 3B

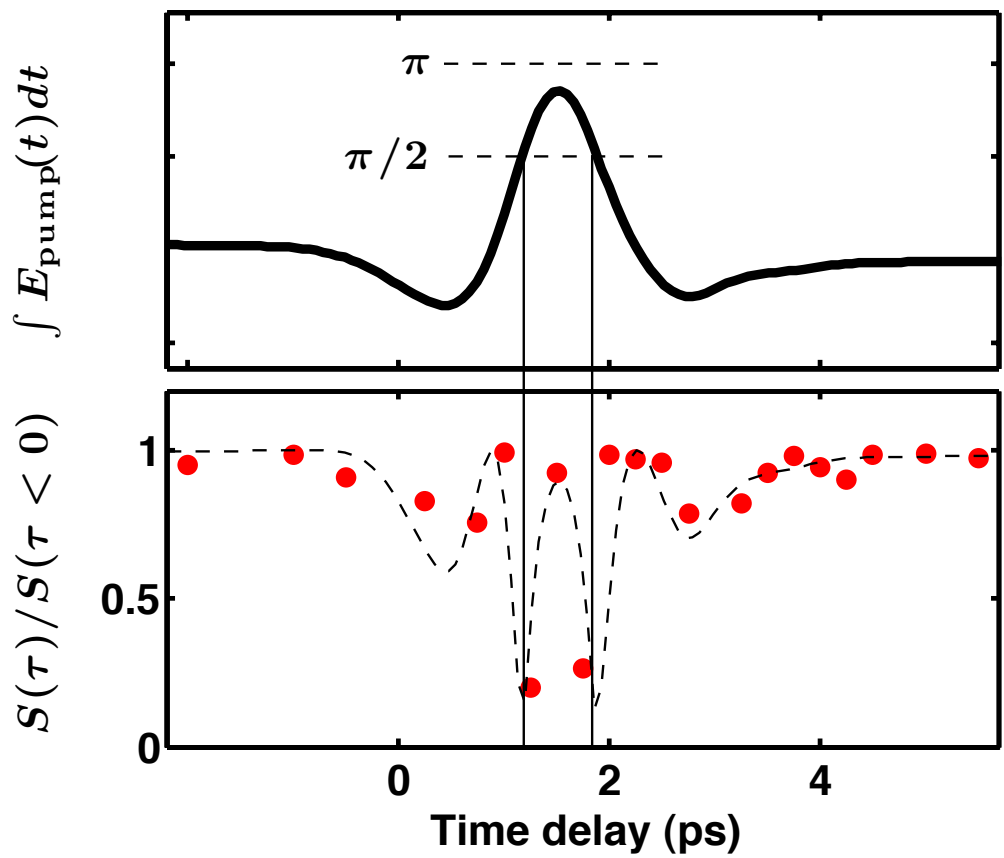

**Figure 4A**

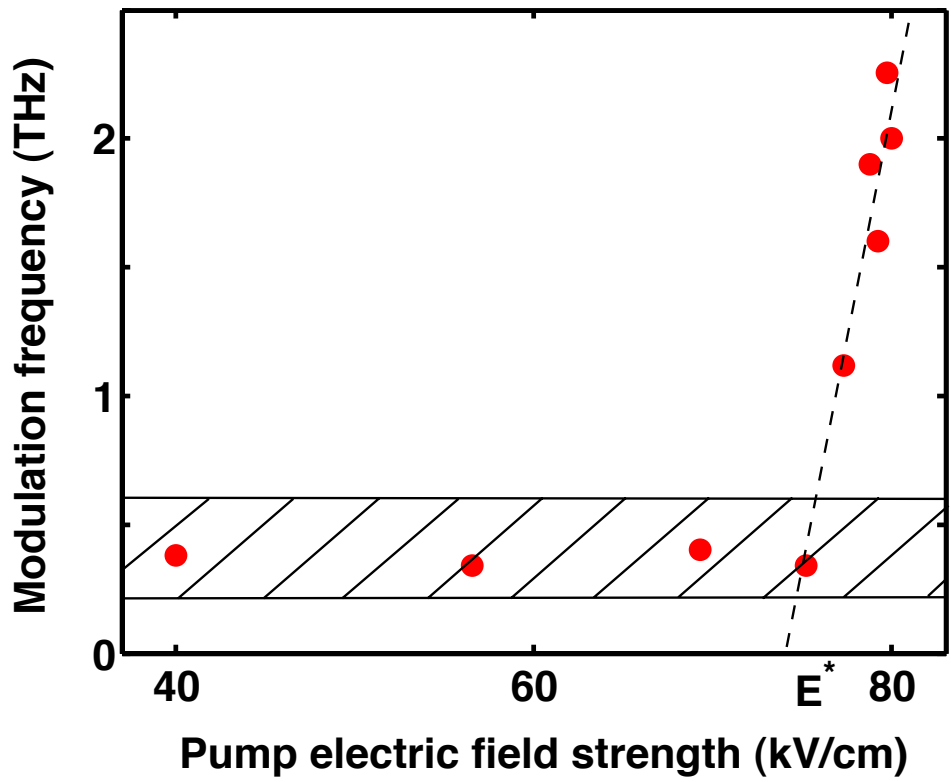

**Figure 4B**

# REFERENCES

*To whom correspondence should be addressed: andreas.dienst@physics.ox.ac.uk or andrea.cavalleri@mpsd.cfel.de

1. Orenstein, J. , Millis A.J., Advances in the Physics of High-Temperature Superconductivity. *Science* **288**, 468 (2000).
2. Tamasaku, K., Nakamura, Y., Uchida, S., Charge dynamics across the $CuO_2$ planes in $La_{2-x}Sr_xCuO_4$. *Phys. Rev. Lett.* **69**, 1455-1458 (1992).
3. Thorsmølle, V. K. *et al.*, C-axis Josephson plasma resonance observed in $Tl_2Ba_2CaCu_2O_8$ superconducting thin films by use of terahertz time-domain spectroscopy. *Optics Letters* **26**, 1292-1294 (2001).
4. Dordevic S. V. *et al.*, Josephson Plasmon and Inhomogeneous Superconducting State in $La_{2-x}Sr_xCuO_4$. *Phys. Rev. Lett.* **91**, 167401 (2003).
5. Basov, D. N. *et al.*, Sum Rules and Interlayer Conductivity of High-$T_c$ Cuprates. *Science* **1**, 49-52 (1999).
6. Anderson, P. W., c-Axis Electrodynamics as Evidence for the Interlayer Theory of High-Temperature Superconductivity. *Science* **20**, 1196-1198 (1998).
7. Moler, K.A. *et al.*, Images of Interlayer Josephson Vortices in $Tl_2Ba_2CuO_{6+\delta}$. *Science* **20**, 1193-1196 (1998).
8. Tsvetkov, A. A. Global and local measures of the intrinsic Josephson coupling in $Tl_2Ba_2CuO_6$ as a test of the interlayer tunnelling model. *Nature* **395**, 360-362 (1998).
9. Schafgans, A. A. *et al.*, Towards a Two-Dimensional Superconducting State of $La_{2-x}Sr_xCuO_4$ in a Moderate External Magnetic Field. *Phys. Rev. Lett.* **104**, 157002 (2010).
10. Kleiner, R. Müller P., Intrinsic Josephson effects in high-$T_c$ superconductors. *Phys. Rev. B* **49**, 1327-1341 (1994).
11. Josephson, B. D. Coupled Superconductors. *Rev. Mod. Phys.* **36**, 216-220 (1964).
12. Josephson, B. D. Possible new effects in superconductive tunneling. *Phys. Lett.* **1**, 251-253 (1962).
13. Hebling J. *et al.*, Generation of high-power terahertz pulses by tilted-pulse-front excitation and their application possibilities. *JOSA B* **25**, B6-B19 (2008).
14. Fausti D. et al. Light induced superconductivity in a striped cuprate. *Science* **331**, 189 (2011).
15. Berg E. *et al.*, Dynamical layer decoupling in a stripe ordered high-$T_c$ superconductor. *Phys. Rev. Lett.* **99**, 127003 (2007).